\documentclass[prd,aps,a4paper,eqsecnum,floatfix,nofootinbib,twocolumn]{revtex4}  %

\usepackage{graphicx} 
\usepackage{amsmath,amsfonts,amssymb}
\usepackage{tikz}
\usepackage{appendix}
\usepackage[caption=false,justification=justified]{subfig}

\newcommand{\beq}{\begin{equation}}
\newcommand{\eeq}{\end{equation}}
\newcommand{\bea}{\begin{eqnarray}}
\newcommand{\eea}{\end{eqnarray}}
\newcommand{\be}{\begin{equation}}      
\newcommand{\ee}{\end{equation}}

\usepackage{slashed}

\begin{document}

\title{Topological Stars and scalar wave equation: Exact resummation of the renormalized angular momentum in the eikonal limit}

\author{Donato Bini$^{1}$, Giorgio Di Russo$^{2}$}
  \affiliation{
$^1$Istituto per le Applicazioni del Calcolo ``M. Picone,''\\ CNR, I-00185 Rome, Italy\\
$^2$School of Fundamental Physics and Mathematical Sciences, Hangzhou Institute for Advanced Study, UCAS, Hangzhou 310024, China\\
}

\date{\today}

\begin{abstract}
We show that for a  Topological Star the renormalized angular momentum parameter, $\nu$, appearing in  the Mano-Suzuki-Takasugi-type or in the quantum-Seiberg-Witten-type approaches of the perturbation equations,  has 1) a direct link with the geodesic radial action computed along the null orbits of the background and 2) admits an exact resummation in terms of hypergeometric functions, generalizing previous results valid in the Schwarzschild case, see Ref. 
[arXiv:2504.07862 [hep-th]].
\end{abstract}

\maketitle
\section{Introduction}
\label{Intro}

When studying the Teukolsky equation for  perturbations (of any spin weight $s$) of the Schwarzschild black hole spacetime the main task is to find explicit solutions of the corresponding homogeneous equation which satisfy the proper boundary conditions: purely ingoing at the horizon (in-solutions) and purely outgoing (up-solutions) at infinity.
This task can be accomplished within some approximation scheme, like Post-Newtonian (PN) approximation, and various procedures are available:
1) MST~\cite{Mano:1996mf,Mano:1996vt} approach, popularized by black hole perturbation reviews  \cite{Sasaki:2003xr,Mino:1997bx} and largely used in the general relativity community;
2) qSW~\cite{Seiberg:1994aj,Seiberg:1994rs,Nekrasov:2002qd,Nekrasov:2009rc,Aminov:2020yma,Bonelli:2022ten,Bianchi:2021mft,Consoli:2022eey} approach, more familiar in the quantum amplitude (and string theory) community.
Recent studies have shown that the two approaches are equivalent, and the computational difficulties they involve (including computational times) are mostly the same, even if in the qSW approach some resummation properties of the PN expansions arise \cite{Bianchi:2024vmi,Cipriani:2025ikx}.

A key feature in both approaches is the expression of the (in, up) solutions of the Teukolsky equation, which is a Confluent Heun equation \cite{heunbook}, as infinite series of hypergeometric functions, with coefficients satisfying a three-term recursion relation (the same relation for both the in and up solutions). The recursion relation, in turn,  implies a compatibility condition which uniquely defines the so-called \lq\lq renormalized angular momentum" function $\nu$ (characterizing, actually, the behavior of the solutions around the irregular singular point at infinity, i.e., the monodromy properties).
Usually, one proceeds perturbatively, and all quantities including $\nu$ are given in a PN-expanded form \cite{Damour:1988mr,Blanchet:2013haa,Bini:2017wfr}, which involves more and more terms depending on the PN accuracy one has planned to reach.

Notice that a basis of two independent solutions to the homogeneous Teukolsky equation can easily be derived: Post-Newtonian-based or Post-Minkowskian-based (or a mixture of both these types), WKB type solutions, ect. The problem (solved by MST and qSW formalisms) is to properly combine them  so that the correct boundary conditions could be satisfied.

Recently, in the Schwarzschild (and Kerr too) case the expression for \lq\lq renormalized angular momentum" $\nu$ has been resummed in terms of hypergeometric functions in the well-defined \lq\lq eikonal limit" \cite{Ivanov:2025ozg}.
This was made possible via a simple relation between the \lq\lq renormalized angular momentum"   and the radial action for (null) geodesics in the same background.
After reviewing the problem (and all recent progress done for its treatment) in the Schwarzschild spacetime, we  show here that this exact resummation can be performed in the case of a Topological Star (TS) \cite{Bah:2020ogh,Bah:2022yji,Bah:2020pdz,Bah:2021irr,Heidmann:2023ojf,Bianchi:2023sfs,DiRusso:2024hmd,Cipriani:2024ygw,Bena:2024hoh,Dima:2024cok,Bianchi:2024vmi,Bianchi:2024rod,Dima:2025zot,DiRusso:2025lip,Bianchi:2025aei,Melis:2025iaw}, the generalization being straightforward but absolutely nontrivial.

The paper is organized as follows: In Sec. II we shortly review the WKB approach to the Teukolsky equation in the Schwarzschild spacetime, showing the existing relation between the eikonal phase, the WKB leading-order solutions and the (null) geodesic radial action~\cite{Kol:2021jjc,Parnachev:2020zbr,Akpinar:2025huz}. 

In Sec. III we discuss the renormalized angular momentum $\nu$ in the Schwarzschild case. Sec. IV generalizes the results of both Sec. II and III to the TS spacetime and contains most of the original contributions of the present work. Conclusions as well as perspectives for future works are then included in Sec. V.

Throughout this paper will use the mostly positive signature of the metric and  units such that $c=1=G$, unless differently specified.

\section{Massless scalar  ($s=0$) wave equation in the Schwarzschild spacetime}

Let us consider the homogeneous massless scalar  ($s=0$) wave equation in the Schwarzschild spacetime,
\beq
\Box\Psi = g^{\mu\nu}\nabla_\mu (\partial_\nu  \Psi)=0\,,
\eeq
for which separation of variables is obtained by looking for solutions with the angular part expressed in terms of (scalar) spherical harmonics: In addition,    Fourier-transforming  the time variable leads to the following (standard) decomposition
\beq
\label{psi_sol}
\Psi(t,r,\theta,\phi)=\sum_{lm} \int \frac{d\omega}{2\pi} e^{-i\omega t} Y_{lm}(\theta,\phi)R_{lm\omega}(r)\,.
\eeq

The radial part of \eqref{psi_sol}, $R_{lm\omega}(r)$, satisfies the homogeneous equation
\bea
\label{eq_phi}
&&\frac{d^2}{dr^2}R_{lm\omega}(r)  + \frac{2 (r-M)}{ r^2f(r)}\frac{d}{dr} R_{lm\omega}(r)\nonumber\\
&&\qquad +\left(\frac{ \epsilon^2}{4M^2 f^2(r)}-\frac{L}{ r^2f(r)}\right) R_{lm\omega}(r)=0\,,
\eea
where, according to standard notation,  $f(r)=1-\frac{2M}{r}$, $\Delta=r^2f(r)$, $L=l(l+1)$, $\epsilon=2M\omega$.

WKB-type solutions to Eq. \eqref{eq_phi} are obtained after replacing
\beq
l=\frac{J}{\hbar}-\frac12\,,\qquad \omega=\frac{E}{\hbar}\,,
\eeq
and following the ansatz
\beq
\label{S_di_r_gen}
R_{lm\omega}(r)= \frac{1}{\sqrt{\Delta}}\,  e^{ i\sum_{k=0}^\infty \hbar^{k-1}S_k(r)}  \,,
\eeq
that is, limiting at the LO and NLO levels, 
\beq
\label{S_di_r}
R_{lm\omega}(r)= \frac{e^{\frac{i}{\hbar}S_0(r)+iS_1(r)}}{\sqrt{\Delta}} +O\left(\hbar\right)\,,
\eeq
with $S_0(r)$ simply related to the null geodesics radial momentum $p_r$
\beq
\label{eq_S_0}
\left(\frac{d S_0(r)}{dr} \right)^2 =\frac{r^3E^2-J^2(r-2M)}{r(r-2M)^2}\equiv p_r^2\,,
\eeq
where, using $u=M/r$ and $j=J/M$ \footnote{When in presence of a massive probe  we will also use the notation $\hat J=J/\mu$ to denote the rescaling of $J$ by the mass of the probe, $\mu$. $\hat J$ has the dimensions of a length while its dimensionless counterpart is $j=\frac{\hat J}{M}=\frac{J}{M\mu}$.}, 
\bea
p_r&=&\frac{\sqrt{r^3E^2-J^2(r-2M)}}{\sqrt{r}(r-2M)}\nonumber\\
&=& \frac{\sqrt{E^2-j^2u^2 (1-2u)}}{(1-2u)}\nonumber\\
&=& \frac{\sqrt{{\mathcal D}}}{(1-2u)}\,,
\eea
with
\beq
{\mathcal D}=E^2-j^2u^2+2j^2u^3=E^2 [1-\hat b^2 u^2(1-2u)]\,. 
\eeq
We also use the notation
\beq
b=\frac{J}{E}\,,\qquad \hat b=\frac{b}{M}\,,
\eeq
the impact parameter so that
\bea
p_r 
&=& E\frac{\sqrt{1-\hat b^2u^2 (1-2u)}}{(1-2u)}\,.
\eea
Finally, as it is well known from WKB approximation
\bea
S_0(r)&=&\pm \int^r p_rdr\,,\nonumber\\
S_1(r)&=&  i \ln (\sqrt{p_r})\,,
\eea
implying
\beq
R_{lm\omega}^\pm(r)= C_\pm \frac{e^{\pm \frac{i}{\hbar} \int^r p_r(r)dr}}{\sqrt{p_r(r)}\sqrt{\Delta}}\,.
\eeq
Choosing $C_\pm=\frac{1}{\sqrt{2}}$ the Wronskian of these solution simplifies as
\bea
W&=& r^2f\left[\left(\frac{d}{dr}R_{lm\omega}^+\right) R_{lm\omega}^- -\left(\frac{d}{dr} R_{lm\omega}^-\right)R_{lm\omega}^+\right]\nonumber\\
&=& \frac{i}{\hbar}\,.
\eea
Working at the leading order (i.e., modulo $O(\hbar)$ terms) is called the {\it eikonal limit}.

The WKB function $S_0(r)$ (solution to Eq. \eqref{eq_S_0}), depending explicitly on $r$ and $b=J/E$, is an approximate WKB solution of Eq. \eqref{eq_phi}.
When considering the limit $r\to \infty$ of $S_0(r)$, Eq. \eqref{S_di_r}, one is looking at the behaviour of the solutions $R_{lm\omega}^\pm(r)$ around the (only) irregular singular point of Eq. \eqref{eq_phi}.  At the same time,  the limit $\lim_{r\to \infty}S(r)$ gives a function of $b$ only, i.e., no more depending on $r$. It coincides with the geodesic radial action (here null geodesic radial action), defined for bound motions as
\bea
\label{I_r_schwt}
I_r^{\rm ell}(b)&=&\frac{1}{2\pi} \oint p_r dr=\frac{1}{\pi}\int_{r_-}^{r_+} p_r dr\,,
\eea
where $r_\pm$ are the periastron and apoastron orbital points.
In other words, the geodesic radial action plays an important role in investigating the behavior of the solutions of the scalar  wave equation (in this specific case, but the treatment can be extended to any spin weight $s$ solutions of the Teukolsky equation), Eq. \eqref{eq_phi}, around its irregular singular point $r=\infty$. 

Since we know that the analytic object responsible for such a behavior is the Floquet index \cite{He:2010if,Bianchi:2021xpr,NISTnu} (or the  renormalized angular momentum parameter, $\nu(l,s=0,\epsilon)$ in the specific case here, present in both approaches MST and qSW), it is expected (and we are going to show this in detail) that  the  Floquet index would  bear  the same information as the geodesic radial action, both  in the eikonal limit where the WKB approximation holds too.

To proceed further, let us  consider the (null) geodesic radial action computed along hyperbolic-like orbits (without any factor of 2, for convenience) defined as
\bea
\label{I_r_schw}
I_r^{\rm hyp}(b)&=&\int_0^{u_{\rm max}} \frac{\sqrt{{\mathcal D}}}{ (1- 2u)} \frac{du}{u^2}\,,
\eea
where $u_{\rm max}\equiv M/r_0$, 
and $r_0$ represents the distance of minimal approach, i.e.,  the biggest root $r_0$ of the cubic equation
\beq
\label{turn}
1-b^2\frac{f(r_0)}{r_0^2}=0\,.
\eeq

Finally, one can connect $I_r^{\rm hyp}(b)$ to $I_r^{\rm ell}(b)$ via the bound-to-unbound map \cite{Kalin:2019rwq,Kalin:2019inp,Cho:2021arx}. 

As it is customary, we will evaluate $I_r^{\rm hyp}(b)$, Eq. \eqref{I_r_schw},  in the large $b$ expansion limit. 
Let us point out since now that, when working in this limit, various variables are commonly used, namely
\bea
x&=& \frac{M}{b}=\frac{1}{\hat b}\ll 1\,,\nonumber\\
y&=& \frac{b_{\rm crit}}{b}=3\sqrt{3}x \ll 1\,, \nonumber\\
z&=& \frac{r_s}{b}= 2 x\ll 1\,.
\eea

In order to evaluate explicitly $I_r(b)$, let us rewrite  the argument of the square root defining $p_r$ along null orbits as
\bea
{\mathcal D}
&=& 2j^2(u-u_1)(u-u_2)(u-u_3)\,.
\eea
In a large-$b$ expansion limit, the three roots result in the following expressions 
\bea
u_1(x) &=& x + x^2    + \frac52 x^3   + 8 x^4 +\frac{231}{8}x^5   +O\left(x^6\right)
 \,,\nonumber\\
u_2(x) &=& u_1(-x)\,, \nonumber\\
u_3(x) &=& \frac12  - 2x^2   - 16 x^4+O\left(x^6\right)\,,
\eea
with $u_{\rm max}=u_1$.  

The exact expression of $r_0=M/u_1$ is the following
\bea
r_0&=&\frac{b^{2/3} \left(\Delta -9M  \right){}^{2/3}+3^{1/3}  b^{4/3}}{3^{2/3} 
(\Delta -9M)^{1/3}}
\,,\nonumber\\
   \Delta&=&  
\sqrt{3}\sqrt{b_{\rm crit}^2-b^2}\,,
\eea
with
\beq
b_{\rm crit}=3\sqrt{3} M\,,  
\eeq
the critical value of the impact parameter implying photons to be captured by the black hole.
One can now proceed to evaluate $I_r^{\rm hyp}(b)$, Eq. \eqref{I_r_schw}, passing to the new variable
\beq
\xi=\frac{u}{u_{\rm max}}\,,
\eeq
and denoting
\bea
\xi_2&=& \frac{u_2}{u_{\rm max}}=-1 +  2 x   - 2 x^2   \nonumber\\
&+& 13 x^3  -24x^4 + \frac{695}{4}x^5+ O\left(x^6 \right) \,,\nonumber\\
\xi_3&=& \frac{u_3}{u_{\rm max}} =- \frac12 +\frac1{2x}   -\frac{11}{4}x \nonumber\\   
&-&  \frac{313}{16}x^3 -\frac{635}{16}x^5+ O\left(x^6\right) \,,
\eea
so that
\bea
I_r^{\rm hyp}(E,b)&=& E b \sqrt{2 u_{\rm max}}\times \nonumber\\
&&  \int_0^1 \frac{d\xi}{\xi^2} \frac{\sqrt{(\xi-1)(\xi-\xi_2)(\xi-\xi_3)}}{(1-2\xi u_{\rm max})}.\qquad
\eea
Expanding  the integrand in large-$b$ immediately gives the sought for result, see below.

One computes then the scattering angle via the well known formula
\beq
\chi=-\frac{\partial I_r^{\rm hyp}}{\partial  J}\,,
\eeq
in a large-$J$ (or large-$b$) expansion limit, obtaining (as a function of $z=\frac{r_s}{b}=\frac{2}{\hat b}=2x$, $\hat b=\frac{b}{M}$)
\bea
\chi\left(z\right)&=&2z  +\frac{15 \pi}{16}z^2+\frac{16}{3}z^3\nonumber\\
   &+&\frac{3465 \pi }{1024}z^4+\frac{112}{5}z^5+\frac{255255 \pi}{16384}z^6\nonumber\\
   &+&\frac{768}{7}z^7+\frac{334639305 \pi 
   }{4194304}z^8+\frac{36608}{63}z^9\nonumber\\
   &+&\frac{29113619535 \pi }{67108864}z^{10}+O\left(z^{11}\right)\,,
   \eea
which exhibits a $\pi$-part (involving even powers of $1/b$) and a non-$\pi$ part (involving odd powers of $1/b$).
Parity (in $b$) properties allow one to isolate then each of these two parts
\bea
\chi_{\pi}\left(z\right)&=&\frac{\chi\left(z\right)+\chi\left(-z\right)}{2}\,,\nonumber\\
\chi_{\slashed{\pi}}\left(z\right)&=&\frac{\chi\left(z\right)-\chi\left(-z\right)}{2}\,,
\eea
namely,
\bea
\chi_{\pi}&=&\pi\left[ \frac{15 }{16}z^2+\frac{3465 }{1024}z^4+\frac{255255 }{16384
   }z^6\right.\nonumber\\
   &+&\left.\frac{334639305  
   }{4194304}z^8+\frac{29113619535  
   }{67108864}z^{10}\right]\nonumber\\
&+& O\left(z^{12}\right)\nonumber\\
\chi_{\slashed{\pi}}&=& 2z +\frac{16}{3}z^3+\frac{112}{5}z^5 +\frac{768}{7}z^7\nonumber\\
   &+&\frac{36608}{63}z^9+O\left(z^{11}\right)\,.
\eea
Passing to the variable $x=\frac{1}{\hat b}$ (also useful for resummation purposes) the above expression  becomes
\bea
\chi\left(x\right)&=&\pi\left[\frac{15}{4} x^2  + \frac{3465}{64} x^4  + \frac{
    255255}{256} x^6  + \frac{334639305}{16384} x^8\right.\nonumber\\  
&+&\left. \frac{29113619535}{65536} x^{10} 
\right]\nonumber\\
&+&
4 x + \frac{128}{3} x^3  +\frac{3584}{5} x^5  + \frac{98304}{7} x^7  \nonumber\\ 
&+&\frac{
 18743296}{63} x^9  +O(x^{11})\,.
\eea

Noticeably, by inspection (i.e., term-by-term factorization) of the various coefficients one can  reconstruct the original function 
   \bea
\label{fullscattsch}
\chi&=&
\frac{1}{2}\sum_{n=1}^\infty c_n y^n \,,
   \eea
with $y=\frac{b_{\rm crit}}{b}=\frac{3\sqrt{3}}{\hat b}$ and 
   where
\beq
c_n=\frac{\Gamma\left(\frac{n}{2}+\frac{1}{6}\right)
\Gamma\left(\frac{n}{2}+\frac{5}{6}\right)}{\Gamma\left(\frac{n}{2}+1\right)^2}\,, 
\eeq
obtaining
\bea\label{SCHscattang}
\chi&=&\frac{2 r_s}{b} \,
   {}_3F_2\left(\frac{2}{3},1,\frac
   {4}{3};\frac{3}{2},\frac{3}{2}
   ;y^2  
 \right) \nonumber\\
&+& \pi  \left(\,
   _2F_1\left(\frac{1}{6},\frac{5
   }{6};1; y^2  
\right)-1\right)\,.
\eea
Finally, the integration over $J$ of Eq. \eqref{fullscattsch} (recalling that $\frac{ME}{J}=\frac{1}{\hat b}=x$) provides us with the radial action
   \beq
I_r^{\rm hyp}=k-\frac{\pi J}{2}+\frac{J y}{4}\sum_{n=1}^\infty\frac{c_n}{n-1}y^{n-1}\,,
   \ee
and $k$ is an integration constant (here $J$ is left explicit for dimensional purposes).
Remarkably, the above result can be expressed in terms of hypergeometric  functions
\bea
I_r^{\rm hyp}&=&k-\frac{J\pi}{2}{}_3F_2\Big[\Big\{-\frac{1}{2},\frac{1}{6},\frac{5}{6}\Big\},\Big\{\frac{1}{2},1\Big\},y^2\Big]\nonumber\\
&+&\frac{32J y^3}{243\sqrt{3}}{}_4F_3\Big[\Big\{1,1,\frac{5}{3},\frac{7}{3}\Big\},\Big\{2,\frac{5}{2},\frac{5}{2}\Big\},y^2\Big]\,.\qquad
\eea
For example, re-expressing $y$ in terms of $x$
\bea
&&1-{}_3F_2\Big[\Big\{-\frac{1}{2},\frac{1}{6},\frac{5}{6}\Big\},\Big\{\frac{1}{2},1\Big\},y^2\Big]=
\frac{15}{4} x^2   \nonumber\\  
&+& \frac{1155}{64} x^4   + \frac{51051}{256}x^6  +\frac{47805615}{16384} x^8 \nonumber\\
&+&  
\frac{3234846615}{65536} x^{10} + O\left(x^{12} \right) \nonumber\\
&=& -2G_0\left(x \right)\,,
\eea
where the function $G_0(x)$ will be introduced below. 
Consequently,
\bea
I_r^{\rm hyp}(b)&=&k-\frac{J\pi}{2}\left(1+2G_0\left(x\right)\right)\nonumber\\
&+&\frac{32J y^3}{243\sqrt{3}}{}_4F_3\Big[\Big\{1,1,\frac{5}{3},\frac{7}{3}\Big\},\Big\{2,\frac{5}{2},\frac{5}{2}\Big\},y^2\Big]\,,\qquad
\eea
with $y=3\sqrt{3}x$ and $x=\frac{1}{\hat b}$.

Note the elliptic-like radial action in the case of a Schwarzschild spacetime was explicitly computed in Refs. \cite{Bini:2020wpo} (see Eq. (13.22)) and \cite{Bini:2020nsb}  (see Eq. (9.5)). 

One can take the eikonal limit ($j\to \infty$ once $\gamma/j=x$ is fixed) of the above mentioned expressions obtaining

\section{The renormalized angular momentum $\nu$ in the Schwarzschild spacetime}

When solving the Teukolsky equation in the Schwarzschild spacetime with proper boundary conditions (and for any spin weight $s$ of the perturbing field) a key role is played by the renormalized angular momentum parameter standardly denoted as $\nu$. One can show that the following expression in powers of $\epsilon=2M\omega$ holds
\bea
\nu(l,s,\epsilon )=l +\sum_{k=1}^\infty \nu_{2k}(l,s)\epsilon^{2k}\,,
\eea
as the compatibility condition of the three-terms recurrence relation
\beq
\alpha_n^\nu a_{n+1}+\beta_n^\nu a_{n}+\gamma_n^\nu a_{n-1}=0\,,
\eeq
with $\alpha_n^\nu, \beta_n^\nu, \gamma_n^\nu$ defined,  for example, in Eq. (124) of Ref. \cite{Sasaki:2003xr} in the Kerr black hole case and hence in the Schwarzschild case (limit $q=0$, $\kappa=1$ according to the notation there).

Recently, Ref. \cite{Ivanov:2025ozg} has shown that the quantity
\beq
\gamma(l, s, \epsilon ) \equiv \nu(l,s, \epsilon )-l
\eeq
is a \lq\lq primitive" object which {\it universally} characterizes a black hole~\footnote{Note that Ref. \cite{Ivanov:2025ozg} expands $\nu$ in powers of $M\omega$ and not $2M\omega$.}. Moreover, Ref. \cite{Ivanov:2025ozg} has proven that $\nu$  is implicitly used to renormalize the multipolar moments of both electric and magnetic type.
Ref.  \cite{Bini:2013rfa}, long ago, pointed out that (in the case $s=-2$ but this is true for any spin weight $s$) the property
\beq
\nu_{2k}(l,-2)+\nu_{2k}(-l-1,-2)=0
\eeq
holds.
One can generalize this relation for any $s$
\beq
\nu_{2k}(l,s)+\nu_{2k}(-l-1,s)=0\,,
\eeq
and then a solution to this equation can be cast in the form
\beq
\nu_{2k}(l,s)=N_{2k}(l,s)-N_{2k}(-l-1,s)\,.
\eeq 
In what follows, we will identify (and study) the functions $N_{2k}(l,s)$ which are the building (fundamental) blocks to construct the universal functions  $\nu(l,s,\epsilon )$ (and have the advantage of involving expressions with half of the terms entering $\nu_{2k}(l,s)$).
Preliminarly, we observe that
\beq
(1+2l)|_{l\to -l-1}=-(1+2l)\,,
\eeq
and, hence, defining
\beq
c(l)=1+2l\,,
\eeq
one finds
\beq
c(l)+c(-l-1)=0\,.
\eeq
Moreover,
\beq
L=l(l+1)|_{l\to -l-1}=l(l+1)=L\,,
\eeq 
namely, $L$ is invariant (and hence any function of $L$ is invariant too), and 
\beq
\frac{dL}{dl}=c(l)\,.
\eeq
Finally, for example, 
\bea
(-1+2l)|_{l\to -l-1}&=& -(3+2l)\,,\nonumber\\
(1+2l)|_{l\to -l-1}&=&-(1+2l)\,,
\eea
and similar relations can be easily found.

To be more explicit, in the case $s=-2$ and $k=1$
\bea
\nu_{2}(l,-2)&=&\left[-\frac{15}{8} + \frac{4}{L} - \frac{225}{8 (-3 + 4 L)}\right]\frac{1}{c(l)}\nonumber\\
&=& 2 N_2(l)\,,
\eea
that is $N_2(l)$ is made of an invariant quantity (depending on $L$) and the term $1/c(l)$, responsible for the sign change as soon as $l\to -l-1$: $N_2(-l-1)=-N_2(l)$.

For $k=2$, instead,
\bea
\nu_{4}(l,-2)&=&\left[  -\frac{1155}{128} - \frac{81}{(-2 + L)} - \frac{8}{L^3} - \frac{54}{L^2} - \frac{42}{L}\right.\nonumber\\
& -& \frac{1225}{
 128 (-15 + 4 L)} - \frac{50625}{8 (-3 + 4 L)^3}\nonumber\\ 
&+&\left. \frac{47325}{
 128 (-3 + 4 L)^2} + \frac{42571}{128 (-3 + 4 L)}\right]\frac{1}{c^3(l)}\nonumber\\
&=& 2 N_4(l)\,,
\eea
and for $k=3$
\bea
\nu_{6}(l,-2)&=& \left[\frac{32}{L^5}+\frac{372}{L^4}+\frac{1469}{L^3}\right.\nonumber\\
&+&\frac{4287}{2 L^2}+\frac{6908}{3 L}-\frac{321489}{524288 (4 L-35)}\nonumber\\
&-&\frac{8506025}{6144 (4 L-15)}-\frac{27783324557}{1572864 (4
   L-3)}\nonumber\\
&-&\frac{370685}{192 (4 L-15)^2}-\frac{604252813}{49152 (4 L-3)^2}\nonumber\\
&+&\frac{7318925}{128 (4 L-3)^3}-\frac{556875}{32 (4 L-3)^4}\nonumber\\
&-&\frac{11390625}{4 (4 L-3)^5}+\frac{3321}{2
   (L-2)}-\frac{8019}{2 (L-2)^2}\nonumber\\
&-&\left. \frac{51051}{512}  \right]\frac{1}{c^5(l)}\nonumber\\
&=& 2 N_6(l)\,.
\eea
Therefore, defining
\beq
\label{G_l_def}
G(l,-2,\epsilon)=\frac{\gamma(l,-2, \epsilon )}{c(l)}\,,
\eeq
one finds
\bea
G(l,-2,\epsilon)&=& A_2(L) \frac{\epsilon^{2}}{c^2(l)} +A_4(L)\frac{\epsilon^{4}}{c(l)^4}\nonumber\\
&+& A_6(L)\frac{\epsilon^{6}}{c(l)^6}+\ldots\nonumber\\
&=& \sum_{k=1}^\infty A_{2k}(L)\left(\frac{\epsilon}{c(l)}\right)^{2k}\nonumber\\
&=& \sum_{k=1}^\infty A_{2k}(L) x^{2k}\,,
\eea
where  we have used the relation
\beq
\frac{\epsilon}{c(l)}= \frac{2M\omega}{2l+1}=\frac{2M\hbar \omega}{2J}=\frac{ME}{J}=\frac{1}{\hat b}=x\,.
\eeq
Following Ref. \cite{Ivanov:2025ozg}, one can consider the limit $L\to \infty$ of the various $A_{2k}(L)$ coefficients at $x$ fixed (eikonal limit, independent of $s$), e.g., 
\bea
\lim_{l\to \infty} A_2(L) &= & -\frac{15 }{8} \,,\nonumber\\
\lim_{l\to \infty} A_4(L) &= &   - \frac{1155}{128} \,,\nonumber\\
\lim_{l\to \infty} A_6(L) &= &  - \frac{51051}{512} \,,
\eea
etc., leading to
\beq
\lim_{l\to \infty}G(l,-2,\epsilon)=G_0(x)\,,
\eeq
where
\bea
G_{(0)}(x)&=&  -\frac{15 x^2}{8}  - \frac{1155}{128} x^4  - \frac{51051}{512} x^6+\ldots.\qquad \nonumber\\
&=&   -\frac{3\cdot 5}{2^3} x^2  - \frac{3\cdot 5 \cdot 7\cdot 11}{2^7} x^4\nonumber\\  
&-& \frac{3\cdot 5 \cdot 7\cdot 11\cdot 13\cdot 17}{2^9} x^6+\ldots.
\eea
The factorization of the various coefficients suggests that a  hypergeometric function can be behind this expansion.
Indeed, as shown in Ref.~\cite{Ivanov:2025ozg}, this expression can be resummed as
\beq
\label{G0x}
G_0(x)=\frac12  \left({}_3F_2\left[\left\{-\frac12, \frac16, \frac56\right\}, \left\{\frac12, 1\right\},27x^2\right] - 1\right)\,,
\eeq 
It follows then the general result that
\bea
I_r^{\rm hyp}&=&k-\frac{J\pi}{2}\left(1+2 \lim_{l\to \infty} G(l,-2,\epsilon)\right)\nonumber\\
&+&\frac{32J y^3}{243\sqrt{3}}{}_4F_3\Big[\Big\{1,1,\frac{5}{3},\frac{7}{3}\Big\},\Big\{2,\frac{5}{2},\frac{5}{2}\Big\},y^2\Big]\,,\qquad
\eea
namely, the $\pi$-part of the radial action is proportional to $I_r^{\rm hyp}$. In turn, this part can be isolated from the full $I_r$ by taking the average
\beq
\frac{I_r^{\rm hyp}(b)+I_r^{\rm hyp}(-b)}{2J}\,,
\eeq
which implies  passing from hyperbolic like orbits to elliptic like orbits via the bound-to-unbound map \footnote{Ref. \cite{Parnachev:2020zbr} has shown that the phase shift in Schwarzschild can be continued analytically for $y>1$ by using the  integral representation
\bea
&&{}_3F_2\left[\{-\frac12, \frac16,\frac56 \},\{\frac12, 1\},y^2  \right]= \nonumber\\
&& -\frac{1}{4\pi}\int_L ds \Gamma(s) \frac{\Gamma(-\frac12 -s)\Gamma(\frac16-s)\Gamma(\frac56-s)}{\Gamma(\frac12-s)\Gamma(1-s)}(-y)^{-s}\nonumber\\
\eea
For $y > 1$, one has  poles at $s = \frac16+ n, \frac56 + n$ and the   contour can be adjusted to include also the pole at $s=\frac12$. Practically, 
the analytic continuation amounts at using proper \lq\lq connection formulae" which allow to express a hypergeomertic function of the variable $y$ as a combination of hypergeometric functions of the variable $1/y$, see Eq. A2 of \cite{Parnachev:2020zbr}.
Clearly, in view of the relation with renormalized angular momentum the same argument applies for $\nu$.}. [Division by $J$ is used to build a dimensionless object.]

Finally, let us display the inverse relation $l=l(\nu)$,
\bea
l&=& \nu+
\frac{ (24 + 13 \nu + 28 \nu^2 + 30 \nu^3 + 15 \nu^4) }{
 2 \nu (1 + \nu) (-1 + 2 \nu) (1 + 2 \nu) (3 + 
    2 \nu) }\epsilon^2\nonumber\\
&+&O(\epsilon^4)\,,
\eea
as well as the product
\bea
l(l+1)&=& \nu (1 + \nu) \nonumber\\
&+& \frac{(24 + 13 \nu + 28 \nu^2 + 30 \nu^3 + 15 \nu^4) }{ 
 2 \nu (1 + \nu) (-1 + 2 \nu) (3 + 2 \nu) }\epsilon^2\nonumber\\ 
&+& O(\epsilon^4)\,.
\eea 

The introduction of the splitting functions $N_{2k}(l,s)$ of $\nu_{2k}(l,s)$ are new with this work. 

We conclude this section by recalling that one cannot use the generic $l$ expressions {\it as they are} for computing the various $\nu_{2k}(l,s)$ when $l$ is specified to some value, e.g. $l=0,1,2,3, \ldots$, unless one corrects them by adding terms of the type $c_2(\delta_2^l-\delta_2^{-l-1})$ for $l=2$, etc. In fact, working with generic $l$ implies multiplying or dividing by factors which may vanish when $l$ is explicitly chosen. This was already pointed out  in Appendix A of Ref.  \cite{Bini:2013rfa} and the reader should refer to this reference for further details.

\section{The case of a topological star}

Let us pass from the familiar  Schwarzschild black hole case to the spacetime of a Topologial Star (TS) \cite{Bah:2020ogh,Bah:2022yji,Bah:2020pdz,Bah:2021irr,Heidmann:2023ojf,Bianchi:2023sfs,DiRusso:2024hmd,Cipriani:2024ygw,Bena:2024hoh,Dima:2024cok,Bianchi:2024vmi,Bianchi:2024rod,Dima:2025zot,DiRusso:2025lip,Bianchi:2025aei,Melis:2025iaw,Bini:2025qyn} which is a solution of Einstein-Maxwell theory in $d=5$
\be
S_5=\int d^5x\sqrt{-g}\left(\frac{R}{2\kappa_5^2}-\frac{1}{2}F_{(m)}^2-\frac{1}{2}F_{(e)}^2\right)
\ee
where $\kappa_5$ is the $d=5$ Einstein gravitational constant, $F_{(m)}$ and $F_{(e)}$ are the magnetic two form and electric three form field strengths. The equations of motion are 
$$
d\star F_{(m)}=d\star F_{(e)}=0
$$
\be
R_{\mu\nu}=\kappa_5^2\left(T_{\mu\nu}-\frac{1}{3}g_{\mu\nu}{T_\alpha}^\alpha\right)
\ee
where $R_{\mu\nu}$ is the Ricci curvature and $T_{\mu\nu}$ is the stress tensor of the gauge fields
\bea
T_{\mu\nu}&{=}&{F_{(m)}}_{\mu\alpha}{{F_{(m)}}_\nu}^\alpha{-}\frac{1}{4}g_{\mu\nu} {F_{(m)}}_{\alpha\beta}{F_{(m)}}^{\alpha\beta}\nonumber\\
&{+}&\frac{1}{2}\Big[{F_{(e)}}_{\mu\alpha\beta}{{F_{(e)}}_\nu}^{\alpha\beta}{-}\frac{1}{6}g_{\mu\nu}{F_{(e)}}_{\alpha\beta\gamma}{F_{(e)}}^{\alpha\beta\gamma}\Big]\,.\qquad
\eea
A spherically symmetric solution for the metric invariant under Wick exchange of $(t,y)$ is
\be
ds^2=-f_s(r) dt^2+\frac{dr^2}{f_s(r) f_b(r)}+r^2d\Omega_2^2+f_b(r) dy^2
\ee
where 
\be
f_s(r)=1-\frac{r_s}{r},\quad f_b(r)=1-\frac{r_b}{r}
\ee
and $d\Omega_2^2$ represents the metric of the 2-sphere.

TS modifications with respect to the Schwarzschild spacetime are summarized by a metric parameter, $\alpha=r_b/r_s$. We refer to the aforementioned papers for more detailed presentation of this geometry. 
We will generalize to this case the discussion of Ref. ~\cite{Ivanov:2025ozg} concerning the resummation of the regularized angular momentum.

\subsection{Radial action for null geodesics in the  spacetime of a Topological Star}

In the spacetime of a TS with fixed $y$ direction, the radial action for null geodesic motion 
\beq
I_r^{\rm hyp}(b) = \int_{r_0}^\infty p_r dr\,,
\eeq
is modified with respect to the Schwarzschild case, because of the parameter $\alpha$ characterizing the metric
\bea
\label{prdef}
p_r 
&=& E\frac{\sqrt{1-\hat b^2u^2 (1-2u)}}{(1-2u)\sqrt{ 1-2\alpha u}}\nonumber\\
&=& \frac{p_r^{\rm Schw}}{\sqrt{ 1-2\alpha u}}\,,
\eea
but the zeros of the numerator do not change with respect to the Schwarzschild case.
Due to this property, the large-$j$ expansion  
proceeds exactly as in the Schwarzschild case with the addition of the $\alpha$-correcting term $( 1-2\alpha u)^{-1/2}$, rewritten as
\beq
( 1-2\alpha u)^{-1/2}=\sum_{k=0}^\infty \frac{\Gamma(1/2)(-2u\alpha)^k }{\Gamma(k + 1)\Gamma(-1/2 - k + 1)}\,,
\eeq 
with $\Gamma(1/2)=\sqrt{\pi}$.

\subsection{The renormalized angular momentum $\nu$ in the  spacetime of a Topological Star}

After having reviewed the properties of the renormalized angular momentum $\nu$ in the familiar Schwarzschild black hole spacetime let us pass to consider the spacetime of a Topologial Star (TS).
When considering   scalar waves ($s=0$) in a TS spacetime we find
\bea
\gamma(l,s=0,\epsilon) &=& \nu(l,s=0,\epsilon)-l\nonumber\\
&=& \nu_2(l,0) \epsilon^2+\nu_4(l,0)\epsilon^4+ \nu_6(l,0)\epsilon^6\nonumber\\
&+&O(\epsilon^8)\,,
\eea
where, for example,
\begin{widetext}
\bea
c(l)\nu_2(l,0) 
&=& -\frac{15}{8} - \frac{1}{8 (-3 + 4 L)} + 
 \alpha \left(-\frac34 + \frac{1}{4 (-3 + 4 L)}\right)+ \alpha^2 \left(-\frac{3}{8} - \frac{1}{8 (-3 + 4 L)} \right) \nonumber\\
c^3(l)\nu_4(l,0) 
&=& -\frac{1155}{128} - \frac{1}{L} - \frac{9}{128 (-15 + 4 L)} - \frac{1}{8 (-3 + 4 L)^3} 
- \frac{483}{ 
 128 (-3 + 4 L)^2} - \frac{949}{128 (-3 + 4 L)}\nonumber\\
&+& 
 \alpha \left(-\frac{105}{32} - \frac{1}{2 L} + \frac{9}{32 (-15 + 4 L)} + \frac{1}{ 
    2 (-3 + 4 L)^3} + \frac{195}{32 (-3 + 4 L)^2} + \frac{265}{32 (-3 + 4 L)}\right)\nonumber\\
&+& 
 \alpha^2 \left(-\frac{105}{64} - \frac{27}{64 (-15 + 4 L)} - \frac{3}{4 (-3 + 4 L)^3} 
- \frac{105}{
    64 (-3 + 4 L)^2} + \frac{57}{64 (-3 + 4 L)}\right)\nonumber\\ 
&+& 
 \alpha^3 \left(-\frac{25}{32} + \frac{9}{32 (-15 + 4 L)} + \frac{1}{2 (-3 + 4 L)^3} + \frac{3}{
    32 (-3 + 4 L)^2} - \frac{31}{32 (-3 + 4 L)}\right)\nonumber\\ 
&+& 
 \alpha^4 \left(-\frac{35}{128} - \frac{9}{128 (-15 + 4 L)} -\frac{ 1}{8 (-3 + 4 L)^3} - \frac{99}{
    128 (-3 + 4 L)^2} - \frac{101}{128 (-3 + 4 L)}\right) \,.
\eea
\end{widetext}
We see that a contribution of the type $\nu_n$ includes powers of $\alpha$ up to $\alpha^n$ (as soon as the $\epsilon$-expansion proceeds), an \lq\lq increasing  modification" of the Schwarzschild spacetime case.
In the TS case the renormalized angular momentum satisfies the three-terms recurrence relation \cite{Bianchi:2024vmi,Bianchi:2024rod,DiRusso:2025lip,Bianchi:2025aei}

\be
\alpha_n^\nu a_{n+1}+\beta_n^\nu a_n+\gamma_n^\nu a_{n-1}=0\,,
\ee
where
\bea
\alpha^\nu_n&{=}&{-}\frac{i \epsilon  \left({\kappa} {+}n{+}\nu {+}1\right) \left({-}{\kappa}{+}n{+}\nu {+}1\right) \left(i {\tau} {+ }n{+}\nu{+}1\right)}{(n{+}\nu {+}1) (2 n{+}2 \nu {+}3)} \nonumber\\
\beta^\nu_n&=&l  (l +1)-(n+\nu ) (n+\nu +1)\nonumber\\
&+&\frac{\epsilon{\kappa}^2 {\tau}}{ (n+\nu ) (n+\nu +1)}\nonumber\\
&-&\frac13(4\tau^2 +  \tau\epsilon +  \epsilon^2) 
\,,\nonumber\\
\gamma^\nu_n&=&\frac{i \epsilon  \left({\kappa}+n+\nu \right) \left(-{\kappa}+n+\nu \right) \left(- i {\tau}+n+\nu
   \right)}{(n+\nu ) (2 n+2 \nu -1)}\,,\nonumber\\
\eea
with
\bea
\label{various_defs}
\epsilon &=&-\omega(r_b-r_s)\,, \nonumber\\
\kappa &=&  \frac{\omega r_s^{3/2}}{(r_b-r_s)^{1/2}}\,,
\nonumber\\
\tau&=&\omega\left(r_s+\frac{r_b}{2}\right)\,.
\eea
The  relations defining $\epsilon$ and $\tau$,  linear in $r_b$ and $r_s$, imply
\beq
\omega r_b=-\frac23 (\epsilon-\tau)\,,\qquad \omega r_s=\frac13 (\epsilon+2\tau)\,,
\eeq
so that  
\beq
27 \kappa^2 \epsilon +(\epsilon+2\tau)^3=0\,.
\eeq
The Schwarzschild limit of Eqs. \eqref{various_defs} corresponds to $r_b=0$.

\subsection{Eikonal limit}
Defining as in Eq. \eqref{G_l_def} (with now $s=0$)
\beq
G_l(0,\epsilon)=\frac{\gamma_l(0, \epsilon )}{c(l)}\,,
\eeq
one writes
\bea
\label{G_l_fun}
G_l(0,\epsilon)
&=& \sum_{k=1}^\infty A_{2k}(L) x^{2k}\,,\qquad x=\frac{\epsilon}{c(l)} 
\,.
\eea
In the eikonal limit, the $\alpha$-modifications to the Schwarzschild values in the coefficients $A_{2k}(L)$ are given by  
\bea
A_2(L)
&=& -\frac{15}{8}    -\frac34 \alpha   -\frac{3}{8}\alpha^2\nonumber\\
A_4(L)
&=& -\frac{1155}{128}  
 -\frac{105}{32} \alpha 
 -\frac{105}{64}\alpha^2  
 -\frac{25}{32}\alpha^3 \nonumber\\ 
& -& \frac{35}{128} \alpha^4  \nonumber\\
A_6(L)&=& -\frac{51051}{512} - \frac{9009}{256} \alpha  - \frac{9009}{512} \alpha^2  - \frac{1155}{128} \alpha^3\nonumber\\  
&-& \frac{2205}{512} \alpha^4  - \frac{441}{256} \alpha^5  - \frac{231}{512} \alpha^6 \nonumber\\
A_8(L)&=& -\frac{47805615}{32768} - \frac{2078505}{4096} \alpha\nonumber\\  
&-& \frac{2078505}{8192} \alpha^2 - \frac{546975}{4096} \alpha^3\nonumber\\  
&-& \frac{1126125}{16384} \alpha^4 
 - \frac{ 135135}{4096} \alpha^5  \nonumber\\ 
&-& \frac{114345}{8192} \alpha^6 -\frac{ 19305}{4096} \alpha^7 - \frac{32175}{32768} \alpha^8 \,.
\eea
Each $A_k(L)$ is a polynom of degree $k$ in $\alpha$.
We can rewrite the above expressions as indicated in Table \ref{tab:1},
\bea
\label{G_l_fun2}
G_l(0,\epsilon)
&=& \sum_{k=1}^\infty A_{2k}^{\alpha^0}(L) x^{2k}
+\alpha \sum_{k=1}^\infty A_{2k}^{\alpha^1}(L) x^{2k} \nonumber\\
&+& \alpha^2 \sum_{k=1}^\infty A_{2k}^{\alpha^2}(L) x^{2k}+\ldots\nonumber\\
&=& G_{(0)}(x)+\alpha G_{(1)}(x)+\alpha^2 G_{(2)}(x)+\ldots
\eea
$G_{(0)}(x)$ was already considered in Eq. \eqref{G0x} above and it is know to be universal, i.e. associated with the Schwarzschild case. The other functions $G_{(i)}(x)$, which characterize a TS, are deviations of order $\alpha^i$ from black hole behaviours. 
Table \ref{tab:1}  below contains their expanded expressions, whereas Table \ref{tab:2} below shows the corresponding resummed forms: these resummations are an original contribution of the present work.

\begin{table*}  
\caption{\label{tab:1}  Expanded form of the functions of the type $G_{(n)}(x)$ ($n=0,\ldots 18$) as defined in Eq. \eqref{G_l_fun}. All expansions are trucated at $O(x^{18})$.
}
\begin{ruledtabular}
\begin{tabular}{ll}
$G_{(0)}(x)$&$ -\frac{15}{8}x^2-\frac{1155}{128}x^4-\frac{51051}{512}x^6 -\frac{47805615}{32768}x^8-\frac{3234846615 x^{10}}{131072}-\frac{957220521075 x^{12}}{2097152}-\frac{75226713509475 x^{14}}{8388608}{-}\frac{395286288806887335 x^{16}}{2147483648}$\\
&$-\frac{33547663138283869575 x^{18}}{8589934592} $\\
$G_{(1)}(x)$&$  -\frac34 x^2-\frac{105}{32}  x^4 - \frac{9009}{256} x^6- \frac{2078505}{4096}x^8-\frac{557732175 x^{10}}{65536}-\frac{82047473235 x^{12}}{524288}-  \frac{12843585233325 x^{14}}{4194304} - \frac{8410346570359305 x^{16}}{134217728}$\\
&$ -\frac{5696772985746317475 x^{18}}{4294967296}$\\
$G_{(2)}(x)$ &$ -\frac{3}{8}x^2-\frac{105}{64}x^4 - \frac{9009}{512}x^6-\frac{2078505}{8192}x^8-\frac{557732175 x^{10}}{131072}{-}\frac{82047473235 x^{12}}{1048576}-\frac{12843585233325 x^{14}}{8388608}-\frac{8410346570359305 x^{16}}{268435456}$\\
&$-\frac{5696772985746317475 x^{18}}{8589934592}$\\
&$ \frac12 G_1(x)$\\
$G_{(3)}(x)$&$  -\frac{25}{32}x^4- \frac{1155}{128}x^6 - \frac{546975}{4096}x^8-\frac{37182145 x^{10}}{16384}-\frac{22055772375 x^{12}}{524288}-\frac{1735619626125 x^{14}}{2097152} - \frac{2281876976454075 x^{16}}{134217728}$\\
&$-\frac{193767788630827125 x^{18}}{536870912}$\\
$G_{(4)}(x)$&$ -\frac{35}{128} x^4  - \frac{2205}{512} x^6  - \frac{1126125}{16384} x^8 -\frac{79214135 x^{10}}{65536}-\frac{47914264125 x^{12}}{2097152}-\frac{3818363177475 x^{14}}{8388608}-\frac{5064653777007825 x^{16}}{536870912}$\\
&$-\frac{432885485239081875 x^{18}}{2147483648}$\\
$G_{(5)}(x)$&$ -\frac{441}{256} x^6  - \frac{135135}{4096} x^8 - \frac{20369349 x^{10}}{32768}-\frac{3194284275 x^{12}}{262144}-\frac{1041371775675 x^{14}}{4194304}-\frac{701259753739545 x^{16}}{134217728}{-}\frac{121207935866942925 x^{18}}{1073741824}$\\
$G_{(6)}(x)$&$ -\frac{231}{512} x^6  - \frac{114345}{8192} x^8 -\frac{19654635 x^{10}}{65536}-\frac{3279465189 x^{12}}{524288}-\frac{1108557051525 x^{14}}{8388608}{-}\frac{764436308130495 x^{16}}{268435456}-\frac{134362285495913475 x^{18}}{2147483648}$\\
$G_{(7)}(x)$&$ -\frac{19305}{4096} x^8 {-} \frac{2147145 x^{10}}{16384}-\frac{794404611 x^{12}}{262144}{-}\frac{70991338275 x^{14}}{1048576}-\frac{202809632769315 x^{16}}{134217728}{-}\frac{18258289666692075 x^{18}}{536870912}$\\
$G_{(8)}(x)$&$ -\frac{32175}{32768} x^8-\frac{6441435 x^{10}}{131072}-\frac{2837159325 x^{12}}{2097152}{-}\frac{276077426625 x^{14}}{8388608}-\frac{829675770419925 x^{16}}{1073741824}{-}\frac{77246610128312625 x^{18}}{4294967296}$\\
$G_{(9)}(x)$&$-\frac{935935 x^{10}}{65536}-\frac{282056775
   x^{12}}{524288}-\frac{62577550035
   x^{14}}{4194304}-\frac{50553720778275
   x^{16}}{134217728}-\frac{19717603185905625
   x^{18}}{2147483648}$\\
$ G_{(10)}(x)$ &$-\frac{323323
   x^{10}}{131072}-\frac{189143955
   x^{12}}{1048576}-\frac{51694497855
   x^{14}}{8388608}-\frac{46369964575935
   x^{16}}{268435456}-\frac{19266915113084925
   x^{18}}{4294967296}$\\
$ G_{(11)}(x)$&$-\frac{24072867
   x^{12}}{524288}-\frac{4699499805
   x^{14}}{2097152}-\frac{9836053091865
   x^{16}}{134217728}-\frac{1114614923897475
   x^{18}}{536870912}$\\
$G_{(12)}(x)$&$-\frac{14196819
   x^{12}}{2097152}-\frac{5688868185
   x^{14}}{8388608}-\frac{15081948074193
   x^{16}}{536870912}-\frac{1929602180080575
   x^{18}}{2147483648}$\\
$G_{(13)}(x)$&$-\frac{643537125
   x^{14}}{4194304}-\frac{1261032447675
   x^{16}}{134217728}-\frac{383873112748125
   x^{18}}{1073741824}$\\
$G_{(14)}(x)$&$-\frac{165480975
   x^{14}}{8388608}-\frac{694854614025
   x^{16}}{268435456}-\frac{274195080534375
   x^{18}}{2147483648}$\\
$G_{(15)}(x)$&$-\frac{70704504585
   x^{16}}{134217728}-\frac{21204419561325
   x^{18}}{536870912}$\\
$G_{(16)}(x)$&$-\frac{128931743655
   x^{16}}{2147483648}-\frac{85739609530575
   x^{18}}{8589934592}$\\
$ G_{(17)}(x)$&$-\frac{7925510124675 x^{18}}{4294967296}$\\
$G_{(18)}(x)$&$-\frac{1622180434875 x^{18}}{8589934592}$\\
\end{tabular}
\end{ruledtabular}
\end{table*}
\begin{table*}  
\caption{\label{tab:2}  Resummed form of the functions of the type $G_{(n)}(x)$ ($n=0,\ldots 6$) as defined in Eq. \eqref{G_l_fun}.
}
\begin{ruledtabular}
\begin{tabular}{ll}
$G_{(0)}(x)$& See Ref. \cite{Ivanov:2025ozg}\\
$G_{(1)}(x)$&$ -\frac34 x^2 {}_3F_2[\{\frac12, \frac56, \frac76\}, \{\frac32, 2\}, 27 x^2] $\\
$G_{(2)}(x)$&$ \frac12 G_1(x)$\\
$G_{(3)}(x)$&$-x^2   \frac54  \left({}_3F_2[\{\frac16, \frac12, \frac56\}, \{1, \frac32\}, 27 x^2] - 
  {}_3F_2[\{\frac16, \frac12, \frac56\}, \{\frac32, 2\}, 27 x^2]\right)  $\\
$G_{(4)}(x)$&$ -x^2  \frac{35}{32} \left(6 {}_3F_2[\{-\frac16, \frac16, \frac12\}, \{1, \frac32\}, 27 x^2] - 
   6 {}_3F_2[\{-\frac16, \frac16, \frac12\},\{\frac32, 2\}, 27 x^2] + 
   x^2 {}_3F_2[\{\frac56, \frac76, \frac32\}, \{2, \frac52\}, 27 x^2]\right)  $\\
$G_{(5)}(x)$&$-x^2 \frac{21}{32} \left(8 {}_3F_2[\{-\frac16, \frac16, \frac12\}, \{1, \frac32\}, 27 x^2] - 
   8 {}_3F_2[\{-\frac16, \frac16, \frac12\}, \{\frac32, 2\}, 27 x^2] + 
   x^2 {}_3F_2[\{\frac56, \frac76, \frac32\}, \{2, \frac52\}, 27 x^2]\right)  $\\
$G_{(6)}(x)$&$-x^2  \frac{77}{331776}   \left(-12 {}_2F_1[-\frac56, \frac56, 2, 27 x^2] + 
   675 x^2 {}_2F_1[\frac16, \frac{11}{6}, 3, 27 x^2] + 
   960 {}_3F_2[\{-\frac16, \frac12, \frac76\}, \{\frac32, 2\}, 27 x^2]\right.$\\ 
&$\left.- 
   1120 {}_3F_2 [\{\frac16, \frac12, \frac56\}, \{\frac32, 2\}, 27 x^2] + 
   172 {}_3F_2 [\{\frac12, \frac56, \frac76\}, \{\frac32, 2\}, 27 x^2]\right) $\\
\end{tabular}
\end{ruledtabular}
\end{table*}

\section{Discussion and concluding remarks}
\label{Conclusion}

We have shown that the renormalized angular momentum parameter, $\nu$, appearing in  the MST-type solutions of the perturbation equations,  can be resummed in terms of simple generalized hypergeometric functions in the case of a Topological Star, generalizing previous results valid in the Schwarzschild case. Our main results are listed in Table \ref{tab:2}, with corresponding expanded form in Table \ref{tab:1}. For example
\bea
G_{(1)}(x)&=&  -\frac34 x^2-\frac{105}{32}  x^4 - \frac{9009}{256} x^6- \frac{2078505}{4096}x^8\nonumber\\
&-&\frac{557732175 x^{10}}{65536}+ O(x^{12}) \nonumber\\
&=& -\frac34 x^2 {}_3F_2[\{\frac12, \frac56, \frac76\}, \{\frac32, 2\}, 27 x^2] \,.
\eea
This nontrivial resummation is a consequence of the relation between $\nu$ and the null geodesic radial action of the background, as already indicated in recent works concerning black holes.  Computational details are summarized in Appendix \ref{ex_resum}.
The importance to have resummed expressions for the renormalized angular momentum $\nu$ is crucial (and motivated the present study too) when one aims at entering the strong field region, which cannot be accessed anyway by using PN expanded expressions (by definition). In fact, the next challenge of black hole perturbation theory is that of reaching accurate modeling of gauge-invariant quantities (like the scattering angle) either beyond the linear order (second-order perturbations seem to be very promising, see e.g., Refs. \cite{Pound:2019lzj,Wardell:2024yoi} for what concerns analytic results) or in the strong field regime.

\section*{Acknowledgments}

We thank J. Parra-Martinez for informative discussions concerning the resummation properties of the renormalized angular momentum in the eikonal limit in the case of black holes and  T. Damour for useful discussions. We also thank G.~Bonelli, A.~Cipriani,  A. Geralico  for comments and suggestions. D.B.  acknowledges sponsorship of
the Italian Gruppo Nazionale per la Fisica Matematica
(GNFM) of the Istituto Nazionale di Alta Matematica
(INDAM). 

\appendix

\section{Top Star and scattering of massive and massless probes}

In this appendix we summarize known results for the scattering angle in the Schwarschild spacetime and in the spacetime of a Topological Star, offering also resummed expressions (when possible) besides the large-$J$ (large-$b$) expanded expressions.

\subsection{Massive probe}
In general, one starts from
 a 4D-reduced TS metric\cite{Bah:2020ogh,Bah:2022yji,Bah:2020pdz,Bah:2021irr,Heidmann:2023ojf,Bianchi:2023sfs,DiRusso:2024hmd,Cipriani:2024ygw,Bena:2024hoh,Dima:2024cok,Bianchi:2024vmi,Bianchi:2024rod,Dima:2025zot,DiRusso:2025lip,Bianchi:2025aei,Melis:2025iaw}
\bea
ds^2|_{4D}&=&-f(r)dt^2+\frac{dr^2}{f(r)\left(1-\frac{2M}{r}\alpha\right)}\nonumber\\
&+& r^2(d\theta^2+\sin^2\theta d\phi^2)\,,
\eea
and writes the mass shell condition in Hamiltonian form for a massive probe as
\beq
\mathcal{H}(x^\alpha,p_\alpha)=-\mu^2\,,
\eeq
yielding the following expression for the scattering angle 
\beq
\label{chi_def}
\chi=\frac{J}{\sqrt{E^2-\mu^2}}\int_{r_1}^\infty \frac{dr}{\sqrt{(r-\alpha r_s)(r-r_1)(r-r_2)(r-r_3)}}\,,
\eeq
where 
\bea
r_1&{=}&\frac{\Delta }{6 \bar{E}}{+}\frac{\frac{r_s^2}{6 \bar{E}}{+}\hat{J}^2}{\Delta }{-}\frac{r_s}{6
   \bar{E}}\,,\nonumber\\
r_2&{=}&{-}\frac{i\left(\sqrt{3}-i\right) \Delta }{12 \bar{E}}{+}\frac{i \left(\sqrt{3}{+}i\right)
   \left(6 \hat{J}^2 \bar{E}{+}r_s^2\right)}{12 \Delta  \bar{E}}{-}\frac{r_s}{6 \bar{E}}\,,\nonumber\\
r_3&{=}&\frac{i \left(\sqrt{3}{+}i\right) \Delta }{12 \bar{E}}{-}\frac{i \left(\sqrt{3}{-}i\right)
   \left(6 \hat{J}^2 \bar{E}{+}r_s^2\right)}{12 \Delta  \bar{E}}{-}\frac{r_s}{6 \bar{E}}\,,
\eea
and we used the notation
\beq
2\bar{E}=\hat{E}^2-1\,,\quad \hat{E}=E/\mu\,,\quad \hat{J}=J/\mu\,,
\eeq
($\mu$ is the mass of the probe) as well as
    \bea
\Delta^3&=&-9 \hat{J}^2 \bar{E} \left(6 \bar{E}+1\right) r_s-r_s^3+3 \sqrt{3} \hat{J} \bar{E}\times \nonumber\\
&&\sqrt{-8
   \hat{J}^4 \bar{E}+\hat{J}^2 \left(36 \bar{E} \left(3 \bar{E}+1\right)-1\right)
   r_s^2+4 r_s^4}\,.\nonumber\\
\eea
$\chi$ is obtained by expanding in large-$\hat J$ the integrand \eqref{chi_def}, and integrating then over $r$ order by order. Its final expression reads
\bea
\label{final_expr}
\chi&=& \sum_{i,j=0}^\infty c_{i,j} \left(\frac{r_s}{\hat{J}}\right)^i\alpha^j\nonumber\\
&=& \chi^{(0)}+\alpha \chi^{(1)}+\alpha^2 \chi^{(2)}+\ldots\,,
\eea
with
\bea
\chi^{(0)}&=& \sum_{i=0}^\infty c_{i,0} \left(\frac{r_s}{\hat{J}}\right)^i\,,\nonumber\\ 
\chi^{(1)}&=& \sum_{i=0}^\infty c_{i,1} \left(\frac{r_s}{\hat{J}}\right)^i\,,\nonumber\\
\chi^{(2)}&=& \sum_{i=0}^\infty c_{i,2} \left(\frac{r_s}{\hat{J}}\right)^i\,,
\eea
etc. The various coefficients $c_{i,j}$ are listed in Table \ref{tab:3} below.
The Schwarzschild limit $\chi^{(0)}$ follows simply by setting $\alpha=0$ in Eq. \eqref{final_expr}. 
Explicitly,
\begin{widetext}    
\bea
\chi^{(0)}&{=}&\frac{\left(4 \bar{E}+1\right) r_s}{\sqrt{2} \hat{J} \sqrt{\bar{E}}}
+\frac{3 \pi  \left(5 \bar{E}+2\right) r_s^2}{8 \hat{J}^2}
+\frac{\left(8 \bar{E} \left(4 \bar{E} \left(16 \bar{E}+9\right)+3\right)-1\right) r_s^3}{24 \sqrt{2} \hat{J}^3 \bar{E}^{3/2}}
+\frac{105 \pi  \left(3 \bar{E} \left(11 \bar{E}+8\right)+4\right) r_s^4}{256 \hat{J}^4}\nonumber\\
   &{+}&\frac{\left(57344 \bar{E}^5{+}51200 \bar{E}^4{+}12800 \bar{E}^3{+}640 \bar{E}^2{-}20
   \bar{E}{+}1\right) r_s^5}{320 \sqrt{2} \hat{J}^5 \bar{E}^{5/2}}
{+}\frac{1155 \pi  \left(221 \bar{E}^3{+}234 \bar{E}^2{+}78 \bar{E}{+}8\right) r_s^6}{2048
   \hat{J}^6}\nonumber\\
   &+&\frac{\left(31457280 \bar{E}^7+38535168 \bar{E}^6+16056320 \bar{E}^5+2508800
   \bar{E}^4+89600 \bar{E}^3-2240 \bar{E}^2+112 \bar{E}-5\right) r_s^7}{17920 \sqrt{2}
   \hat{J}^7 \bar{E}^{7/2}}\nonumber\\
   &+&\frac{45045 \pi  \left(7429 \bar{E}^4+10336 \bar{E}^3+5168 \bar{E}^2+1088
   \bar{E}+80\right) r_s^8}{262144 \hat{J}^8}\nonumber\\
   &+&\frac{2r_s^9}{258048 \sqrt{2} \hat{J}^9 \bar{E}^{9/2}}\Big[4798283776 \bar{E}^9{+}7474249728 \bar{E}^8{+}4359979008 \bar{E}^7{+}1156055040
   \bar{E}^6{+}130056192 \bar{E}^5\nonumber\\
   &+&3612672 \bar{E}^4-75264 \bar{E}^3+3456 \bar{E}^2-180
   \bar{E}{+}7 \Big]\nonumber\\
   &+&\frac{2909907 \pi  \left(10005 \bar{E}^5+17250 \bar{E}^4+11500 \bar{E}^3+3680
   \bar{E}^2+560 \bar{E}+32\right) r_s^{10}}{2097152 \hat{J}^{10}}\nonumber\\
   &+&+\frac{r_s^{11}}{1081344 \sqrt{2} \hat{J}^{11} \bar{E}^{11/2}}\Big[223338299392 \bar{E}^{11}+422248972288 \bar{E}^{10}+316686729216
   \bar{E}^9+118757523456 \bar{E}^8\nonumber\\
   &+&22837985280 \bar{E}^7+1998323712 \bar{E}^6+45416448
   \bar{E}^5-811008 \bar{E}^4+33792 \bar{E}^3-1760 \bar{E}^2+88 \bar{E}-3
   \Big]\nonumber\\
   &+&\frac{22309287 \pi  \left(471975 \bar{E}^6+970920 \bar{E}^5+809100 \bar{E}^4+348000
   \bar{E}^3+81000 \bar{E}^2+9600 \bar{E}+448\right) r_s^{12}}{67108864 \hat{J}^{12}}\nonumber\\
   &+&\frac{r_s^{13}}{112459776
   \sqrt{2} \hat{J}^{13} \bar{E}^{13/2}}\Big[266356691828736 \bar{E}^{13}+592293169987584 \bar{E}^{12}+542935405821952
   \bar{E}^{11}\nonumber\\
   &+&263483358707712 \bar{E}^{10}+72046230896640 \bar{E}^9+10806934634496
   \bar{E}^8+771923902464 \bar{E}^7\nonumber\\
   &+&14844690432 \bar{E}^6-231948288 \bar{E}^5+8785920
   \bar{E}^4-439296 \bar{E}^3+23296 \bar{E}^2-1092 \bar{E}+33\Big]+O\left(\frac{r_s}{\hat J}\right)^{14}\,.
\eea
\end{widetext}
 
\begin{table*}  
\caption{\label{tab:3} TS and massive probe: list of the various coefficients entering the scattering angle. Noticeably the $\alpha \to0$ limit of the scattering angle reproduces the corresponding Schwarzschild values.
}
\begin{ruledtabular}
\begin{tabular}{ll}
$c_{1,0}$& $\frac{4 \bar{E}+1}{\sqrt{2} \sqrt{\bar{E}}}$\\ 
$c_{1,1}$& $\sqrt{2} \sqrt{\bar{E}}$\\
\hline
$c_{2,0}$& $\frac{3}{8} \pi  \left(5 \bar{E}+2\right)$\\ 
$c_{2,1}$& $\frac{1}{4} \pi  \left(3 \bar{E}+1\right)$\\ 
$c_{2,2}$& $\frac{3 \pi  \bar{E}}{8}$\\
\hline
$c_{3,0}$& $\frac{512 \bar{E}^3+288 \bar{E}^2+24 \bar{E}-1}{24 \sqrt{2} \bar{E}^{3/2}}$\\ 
$c_{3,1}$& $\frac{32 \bar{E}^2+16 \bar{E}+1}{4 \sqrt{2} \sqrt{\bar{E}}}$\\ 
$c_{3,2}$& $ \frac{\sqrt{\bar{E}} \left(8 \bar{E}+3\right)}{2 \sqrt{2}}$\\ 
$c_{3,3}$&$\frac{5 \bar{E}^{3/2}}{3 \sqrt{2}}$\\
\hline
$c_{4,0}$& $\frac{105}{256} \pi  \left(33 \bar{E}^2{+}24 \bar{E}{+}4\right)$\\ 
$c_{4,1}$& $\frac{15}{64} \pi  \left(21 \bar{E}^2{+}14 \bar{E}{+}2\right)$\\ 
$c_{4,2}$& $\frac{9}{128} \pi  \left(35 \bar{E}^2{+}20 \bar{E}{+}2\right)$\\
$c_{4,3}$& $ \frac{15}{64} \pi  \bar{E} \left(5 \bar{E}{+}2\right)$\\
$c_{4,4}$& $ \frac{105 \pi  \bar{E}^2}{256}$\\
\hline
$c_{5,0}$& $\frac{57344 \bar{E}^5+51200 \bar{E}^4+12800 \bar{E}^3+640 \bar{E}^2-20 \bar{E}+1}{320
   \sqrt{2} \bar{E}^{5/2}}$\\
$c_{5,1}$& $ \frac{6144 \bar{E}^4+5120 \bar{E}^3+1152 \bar{E}^2+48 \bar{E}-1}{96 \sqrt{2}
   \bar{E}^{3/2}}$\\
$c_{5,2}$& $\frac{\left(4 \bar{E}{+}1\right) \left(64 \bar{E}^2{+}32 \bar{E}{+}1\right)}{8 \sqrt{2}
   \sqrt{\bar{E}}}$\\
$c_{5,3}$& $\frac{\sqrt{\bar{E}} \left(64 \bar{E}^2{+}40 \bar{E}{+}5\right)}{4 \sqrt{2}}$\\
$c_{5,4}$& $\frac{7 \bar{E}^{3/2} \left(12 \bar{E}{+}5\right)}{12 \sqrt{2}}$\\
$c_{5,5}$& $\frac{21 \bar{E}^{5/2}}{10 \sqrt{2}}$\\
\hline
$c_{6,0}$& $\frac{1155 \pi  \left(221 \bar{E}^3+234 \bar{E}^2+78 \bar{E}+8\right)}{2048}$\\
$c_{6,1}$& $\frac{315 \pi  \left(143 \bar{E}^3+143 \bar{E}^2+44 \bar{E}+4\right)}{1024}$\\
$c_{6,2}$& $\frac{105 \pi  \left(429 \bar{E}^3+396 \bar{E}^2+108 \bar{E}+8\right)}{2048}$\\
$c_{6,3}$& $\frac{25}{512} \pi  \left(231 \bar{E}^3+189 \bar{E}^2+42 \bar{E}+2\right)$\\
$c_{6,4}$& $ \frac{525 \pi  \bar{E} \left(21 \bar{E}^2+14 \bar{E}+2\right)}{2048}$\\
$c_{6,5}$& $\frac{315 \pi  \bar{E}^2 \left(7 \bar{E}+3\right)}{1024}$\\
$c_{6,6}$& $\frac{1155 \pi  \bar{E}^3}{2048}$\\
\hline
$c_{7,0}$& $\frac{31457280 \bar{E}^7+38535168 \bar{E}^6+16056320 \bar{E}^5+2508800 \bar{E}^4+89600
   \bar{E}^3-2240 \bar{E}^2+112 \bar{E}-5}{17920 \sqrt{2} \bar{E}^{7/2}}$\\
$c_{7,1}$& $\frac{786432 \bar{E}^6+917504 \bar{E}^5+358400 \bar{E}^4+51200 \bar{E}^3+1600
   \bar{E}^2-32 \bar{E}+1}{1280 \sqrt{2} \bar{E}^{5/2}}$\\
$c_{7,2}$& $\frac{3 \left(65536 \bar{E}^5+71680 \bar{E}^4+25600 \bar{E}^3+3200 \bar{E}^2+80
   \bar{E}-1\right)}{640 \sqrt{2} \bar{E}^{3/2}}$\\
$c_{7,3}$& $\frac{5 \left(2048 \bar{E}^4+2048 \bar{E}^3+640 \bar{E}^2+64 \bar{E}+1\right)}{64
   \sqrt{2} \sqrt{\bar{E}}}$\\
$c_{7,4}$& $\frac{5 \sqrt{\bar{E}} \left(512 \bar{E}^3{+}448 \bar{E}^2{+}112 \bar{E}{+}7\right)}{32
   \sqrt{2}}$\\
$c_{7,5}$& $\frac{9 \bar{E}^{3/2} \left(320 \bar{E}^2{+}224 \bar{E}{+}35\right)}{80 \sqrt{2}}$\\
$c_{7,6}$& $\frac{33 \bar{E}^{5/2} \left(16 \bar{E}{+}7\right)}{40 \sqrt{2}}$\\
$c_{7,7}$& $\frac{429 \bar{E}^{7/2}}{140 \sqrt{2}}$\\
\hline
$c_{8,0}$& $\frac{45045 \pi  \left(7429 \bar{E}^4{+}10336 \bar{E}^3{+}5168 \bar{E}^2{+}1088
   \bar{E}{+}80\right)}{262144}$\\ 
$c_{8,1}$& $\frac{15015 \pi  \left(969 \bar{E}^4{+}1292 \bar{E}^3{+}612 \bar{E}^2{+}120
   \bar{E}{+}8\right)}{32768}$\\
$c_{8,2}$& $\frac{3465 \pi  \left(4199 \bar{E}^4{+}5304 \bar{E}^3{+}2340 \bar{E}^2{+}416
   \bar{E}{+}24\right)}{65536}$\\
$c_{8,3}$& $ \frac{1575 \pi  \left(2431 \bar{E}^4{+}2860 \bar{E}^3{+}1144 \bar{E}^2{+}176
   \bar{E}{+}8\right)}{32768}$\\
$c_{8,4}$& $\frac{1225 \pi  \left(6435 \bar{E}^4+6864 \bar{E}^3+2376 \bar{E}^2+288
   \bar{E}+8\right)}{131072}$\\
$c_{8,5}$& $\frac{2205 \pi  \bar{E} \left(429 \bar{E}^3+396 \bar{E}^2+108 \bar{E}+8\right)}{32768}$\\
$c_{8,6}$& $\frac{24255 \pi  \bar{E}^2 \left(33 \bar{E}^2+24 \bar{E}+4\right)}{65536}$\\
$c_{8,7}$& $\frac{15015 \pi  \bar{E}^3 \left(9 \bar{E}+4\right)}{32768}$\\
$c_{8,8}$& $\frac{225225 \pi  \bar{E}^4}{262144}$\\
\end{tabular}
\end{ruledtabular}
\end{table*}

\subsection{TS: massless probe}

In the case of a TS spacetime and a massless probe (expressing the results in powers of $z=r_s/b$) we find

\bea
\chi_0^{(0)}&{=}&\sum_{i}c_{i,0}z^i\nonumber\\
&=&2z{+}\frac{15 \pi}{16}z^2 {+}\frac{16}{3}z^3{+}\frac{3465 \pi}{1024}z^4\nonumber\\
&{+}&\frac{112}{5}z^5{+}\frac{255255 \pi}{16384}z^6{+}\frac{768}{7}z^7{+}\frac{334639305 \pi}{4194304}z^8\nonumber\\
&+&\frac{36608}{63}z^9+\frac{29113619535 \pi}{67108864}z^{10}+\frac{106496}{33}z^{11}\nonumber\\
&{+}&\frac{10529425731825 \pi}{4294967296}z^{12}{+}\frac{2646016}{143}z^{13}\nonumber\\
&{+}&O\left(z^{14}\right)\,,
\eea
Note that the expression $\chi_0^{(0)}$ is called $\phi_{\rm M}$ in Ref. \cite{Parnachev:2020zbr} (see Eqs. (3.5) and (3.6), $\phi_{\rm M}$ there coincides with $\chi_0^{(0)}$ here)
and the following expression is also given (in $D=4$)
\beq
\chi_0^{(0)}=\sum_{k=1}^\infty \frac{\sqrt{\pi}}{k!}\frac{\Gamma \left(\frac{3k+1}{2}\right)}{\Gamma \left(\frac{k}{2}+1\right)}z^k\,,
\eeq
which coincides with Eq. \eqref{SCHscattang} above.
Moreover,
\bea
\chi_0^{(1)}&=&\sum_{i}c_{i,1}z^i\nonumber\\
&=&z+\frac{3 \pi}{8}z^2+2z^3 \nonumber\\
&{+}&\frac{315 \pi}{256}z^4{+}8z^5 {+}\frac{45045 \pi }{8192}z^6{+}\frac{192}{5}z^7\nonumber\\
&{+}&\frac{14549535 \pi}{524288}z^8{+}\frac{1408}{7}z^9{+}\frac{5019589575 \pi}{33554432}z^{10}\nonumber\\
&{+}&\frac{3328}{3}z^{11}+\frac{902522205585 \pi}{1073741824}z^{12}\nonumber\\
&+&\frac{69632}{11}z^{13}+O\left(z^{14}\right)\,,
\eea
\bea
\chi_0^{(2)}&=&\sum_{i}c_{i,2}z^i\nonumber\\
&=&\frac{3 \pi}{16}z^2+z^3+\frac{315 \pi}{512}z^4+4z^5\nonumber\\
&+& \frac{45045 \pi}{16384}z^6+\frac{96}{5}z^7\nonumber\\
&{+}&\frac{14549535 \pi}{1048576}z^8{+}\frac{704}{7}z^9{+}\frac{5019589575 \pi}{67108864}z^{10}\nonumber\\
&{+}&\frac{1664}{3}z^{11}{+}\frac{902522205585 \pi}{2147483648}z^{12}\nonumber\\
&{+}&\frac{34816}{11} z^{13}+O\left(z^{14}\right)\,,
\eea
and
\bea
\chi_0^{(3)}&=&\sum_{i}c_{i,3}z^i\nonumber\\
&=&\frac{5}{12}z^3+\frac{75 \pi }{256}z^4
+2z^5 +\frac{5775 \pi}{4096}z^6+10z^7 \nonumber\\
&{+}&\frac{3828825 \pi }{524288}z^8{+}\frac{160}{3}z^9{+}\frac{334639305 \pi }{8388608} z^{10}\nonumber\\
&{+}&\frac{2080}{7}z^{11}+\frac{242613496125 \pi}{1073741824} z^{12}\nonumber\\
&+&\frac{5120}{3} z^{13}+O\left(z^{14}\right)\,.
\eea

The previous expansions can be resummed (another original contribution with this work) 
\bea
\chi_0^{(1)}&{=}&z \, _3F_2\left(\frac{1}{3},\frac{2}{3},1;\frac{1}{2},\frac{3}{2};\frac{27
   }{4}z^2\right) \nonumber\\
   &+&\frac{3 \pi}{8}z^2 \,
   _2F_1\left(\frac{5}{6},\frac{7}{6};2;\frac{27}{4}z^2\right)\,,
   \eea
   \bea
\chi_0^{(2)}&{=}&z^2 \, _3F_2\left(1,\frac{4}{3},\frac{5}{3};\frac{3}{2},\frac{5}{2};\frac{27
  }{4}z^2\right) \nonumber\\
   &+&\frac{3 \pi }{16}z \left( \,
   _2F_1\left(\frac{5}{6},\frac{7}{6};2;\frac{27}{4}z^2\right)-1\right)\,,
   \eea
and
    \bea
\chi_0^{(3)}&{=}&\frac{5 }{12  }z\,\left( _3F_2\left(\frac{2}{3},1,\frac{4}{3};\frac{1}{2},\frac{5}{2};\frac{27
   }{4}z^2\right)-1\right)\nonumber\\
   &+&\frac{75 \pi}{256}z^2 \,
   \left(_2F_1\left(\frac{7}{6},\frac{11}{6};3;\frac{27}{4}z^2\right)-1\right)\,,\qquad
   \eea
while $\chi_0^{(0)}$ appears in \eqref{SCHscattang}.

\section{Exact resummation of TS radial action: short review of computational details}
\label{ex_resum}

Starting from the radial momentum of massless probes in TS geometry
\be
p_r^2=\frac{E^2r^3- J^2(r-r_s)}{(r-r_s)^2(r-\alpha r_s)}\,,
\ee
so that
\bea
I_r&=&\int_{r_0}^\infty p_rdr\nonumber\\ 
&=&\int_0^{u_{\rm max}}  \frac{ E M\sqrt{1 - \hat b^2 u^2 (1-2u)}}{(1 - 2 u) u^2 \sqrt{1 - 2 \alpha u}}du\,,  
\eea
where the minimum approach distance $r_0$ is such that $M/r_0=u_{\rm max}$.
Changing the variable 
\beq
u=\frac{\xi}{\hat b}
\eeq
we find
\bea
I_r 
&=&\int_0^{\hat b u_{\rm max}}  \frac{ M E\sqrt{1 - \xi^2 (1-2\frac{\xi}{\hat b})}}{(1 - 2 \frac{\xi}{\hat b}) \frac{\xi^2}{\hat b^2}  \sqrt{1 - 2 \alpha \frac{\xi}{\hat b}}}\frac{d\xi}{\hat b}\nonumber\\
&=&E M \hat b \int_0^{\hat b u_{\rm max}} \frac{\sqrt{1-\xi^2}}{ \xi^2   } {\mathcal A}(\xi, \frac{1}{\hat b}) d\xi\,,
\eea
where
\beq
{\mathcal A}(\xi, \frac{1}{\hat b})= \frac{\sqrt{1+2\frac{\xi^3}{(1-\xi^2)\hat b}}}{(1 - 2 \frac{\xi}{\hat b} ) \sqrt{1 - 2 \alpha \frac{\xi}{\hat b}}}
\eeq
also
recalling that
\bea
\hat b u_{\rm max}&=& 1+ \frac{1}{\hat b}+ \frac{5}{2 \hat b^2}+O\left(\frac{1}{\hat b^3}\right)\,.  
\eea
We can now expand in large-$\hat b$ the term ${\mathcal A}(\xi, \frac{1}{\hat b})$ 
so that
\bea
&&\mathcal{A}\left(\xi, \frac{1}{\hat{b}}\right) 
= -\sum_{n,k=0}^\infty \frac{2^{n-1}}{\sqrt{\pi} \, n! \, \hat{b}^n} 
\left(-\frac{\xi^3}{1 - \xi^2}\right)^n 
\Gamma\left(n - \frac{1}{2}\right) \nonumber\\
&\times& {}_2F_1\left(1, -n; \frac{3}{2} - n; -\frac{1 - \xi^2}{\xi^2} \right)\binom{-\frac{1}{2}}{k}\left({-}\frac{2\alpha\xi}{\hat{b}}\right)^k\,.
\eea
In order to integrate the previous expression it is useful to write the hypergeometric function in its series representation
\bea
&&{}_2F_1\left(1, -n; \frac{3}{2} - n; -\frac{1 - \xi^2}{\xi^2} \right)=\nonumber\\
&&\sum_{s=0}^\infty \frac{\Gamma
   \left(\frac{3}{2}-n\right)
    \Gamma
   (s-n)}{\Gamma (-n) \Gamma
   \left(-n+s+\frac{3}{2}\right)}\left(1-\frac{1}{\xi
   ^2}\right)^s\,.
\eea
Let us define
\be
I_r(\hat b)=\sum_{n,k,s=0}^\infty \mathcal{I}_{nks}(\hat{b})\,,
\ee
where 
\begin{widetext}    
\beq
\mathcal{I}_{nks}(\hat{b})=\frac{E M
   \binom{-\frac{1}{2}}{k}
   \alpha ^k 2^{k+n-1} \tan (\pi
    n) \hat{b}^{-k-n+1}
   (-1)^{k+n+s} \Gamma
   (s-n)}{\sqrt{\pi } \Gamma
   \left(-n+s+\frac{3}{2}\right)
   }\int_0^{\hat{b}u_{\rm max}}d\xi \left(1-\xi
   ^2\right)^{-n+s+\frac{1}{2}}
   \xi ^{k+3 n-2 s-2}\,.
\eeq
It is enough to consider  the leading-order term in the PM expansion of the upper integration limit $\hat b u_{\rm max}=1+O(1/{\hat b})$. Consequently 
\beq
\mathcal{I}_{nks}(\hat{b})=\frac{E M
   \binom{-\frac{1}{2}}{k}
   \alpha ^k 2^{k+n-2} \tan (\pi
    n) \hat{b}^{-k-n+1}
   (-1)^{k+n+s} \Gamma (s-n)
   \Gamma \left(\frac{1}{2} (k+3
   n-2 s-1)\right)}{\sqrt{\pi }
   \Gamma \left(\frac{1}{2}
   (k+n+2)\right)}\,.
\eeq
The sum over $s$ yields
\beq
\mathcal{I}_{nk}(\hat{b})=\sum_{s=0}^\infty \mathcal{I}_{nks}(\hat{b})=\frac{E (-1)^k M e^{i \pi  n}
   \binom{-\frac{1}{2}}{k}
   \alpha ^k 2^{k+n-1} \tan (\pi
    n) \Gamma (-n)
   \hat{b}^{-k-n+1} \Gamma
   \left(\frac{1}{2} (k+3
   n+1)\right)}{\sqrt{\pi }
   (k+n-1) \Gamma
   \left(\frac{1}{2}
   (k+n+2)\right)}\,,
\eeq
while that over $k$ gives
\bea
\mathcal{I}_{n}(\hat{b})&=&\sum_{k=0}^\infty \mathcal{I}_{nk}(\hat{b})=\frac{1}{8} \pi ^{3/2} E M e^{i \pi  n} \hat{b}^{-n} \sec (\pi  n)
   \Bigg[-\frac{3 \alpha  \Gamma \left(\frac{3 n}{2}\right) \,
   _4\tilde{F}_3\left(\frac{3}{4},\frac{5}{4},\frac{n}{2},\frac{3
   n}{2}+1;\frac{3}{2},\frac{n+2}{2},\frac{n+3}{2};\frac{4 \alpha
   ^2}{\hat{b}^2}\right)}{\Gamma \left(\frac{n+1}{2}\right)}\nonumber\\
   &-&\frac{4 \hat{b}
   \Gamma \left(\frac{1}{2} (3 n+1)\right) \,
   _4\tilde{F}_3\left(\frac{1}{4},\frac{3}{4},\frac{n-1}{2},\frac{1}{2} (3
   n+1);\frac{1}{2},\frac{n+1}{2},\frac{n+2}{2};\frac{4 \alpha
   ^2}{\hat{b}^2}\right)}{(n-1) \Gamma \left(\frac{n}{2}+1\right)}\Bigg]\,.
\eea
The latter can be then expanded in powers of $\alpha$, and each order can be resummed separately
\bea
I_r^{\alpha^0}&=&-\frac{E M \sqrt{\pi }}{2}\sum_{n=0}^\infty\frac{  2^{n}
   e^{i \pi  n} \hat{b}^{1-n}
   \sec (\pi  n) \Gamma
   \left(\frac{1}{2} (3
   n+1)\right)}{(n-1) \Gamma
   (n+1) \Gamma
   \left(\frac{n+2}{2}\right)}\nonumber\\
   &=&\frac{E M \left(3 \pi  \hat{b}^3
   \,
   _3F_2\left(-\frac{1}{2},\frac{1}{6},\frac{5}{6};\frac{1}{2},1;\frac{27}{\hat{b}^2}\right)-64 \,
   _4F_3\left(1,1,\frac{5}{3},\frac{7}{3};2,\frac{5}{2},\frac{5}{2};\frac{27}{\hat{b}^2}\right)\right)}{6 \hat{b}^2}\nonumber\\
I_r^{\alpha^1}&=&-\frac{3 i E M \sqrt{\pi }}{4}\sum_{n=0}^\infty \frac{ 
   2^{n} \hat{b}^{-n} (\tan
   (\pi  n)-i) \Gamma
   \left(\frac{3
   n}{2}\right)}{\Gamma (n+1)
   \Gamma
   \left(\frac{n+3}{2}\right)}\nonumber\\
   &=&
-\frac{E M \left(16 \,
   _4F_3\left(1,1,\frac{4}{3},\frac{5}{3};\frac{3}{2},2,\frac
   {5}{2};\frac{27}{\hat{b}^2}\right)+3 \pi  \hat{b} \,
   _3F_2\left(\frac{1}{2},\frac{
   5}{6},\frac{7}{6};\frac{3}{2}
   ,2;\frac{27}{\hat{b}^2}\right
   )\right)}{4 \hat{b}^2}\nonumber\\
   I_r^{\alpha^2}&=&-\frac{3 i E M \sqrt{\pi }}{\hat{b}}\sum_{n=0}^\infty \frac{2^n \hat{b}^{-n} (\tan
   (\pi  n)-i) \Gamma
   \left(\frac{3
   (n+1)}{2}\right)}{n \Gamma
   \left(\frac{n}{2}\right)
   \Gamma (n+3)}\nonumber\\
   &=&\frac{E M \left(105 \pi  \,
   _3F_2\left(\frac{3}{2},\frac{
   11}{6},\frac{13}{6};\frac{5}{
   2},3;\frac{27}{\hat{b}^2}\right)-4 \hat{b} \left(16 \,
   _4F_3\left(1,1,\frac{4}{3},\frac{5}{3};\frac{3}{2},2,\frac
   {5}{2};\frac{27}{\hat{b}^2}\right)+3 \pi  \hat{b} \,
   _2F_1\left(\frac{5}{6},\frac{
   7}{6};2;\frac{27}{\hat{b}^2}\right)\right)\right)}{32
   \hat{b}^3}\,,
 \eea
 where in general each $I_r^{\alpha^i}$ is defined up to a constant term which 
accounts for UV divergences. 
\end{widetext}

\end{document}